\documentclass[aps,pra,twocolumn,superscriptaddress,reprint]{revtex4-1}

\usepackage{float}
\usepackage{graphicx}
\usepackage{amsmath}
\usepackage{bm}
\usepackage{color}
\usepackage{amssymb}
\usepackage{mathrsfs}
\usepackage{dcolumn}
\usepackage{multirow}
\usepackage[colorlinks=true,linkcolor=blue,urlcolor=blue,citecolor=blue]{hyperref}

\begin{document}
	
	\title{Simulating the Palmer-Chalker state in an orbital superfluid}
	
	\author{Hua Chen}
	\email{Electronic address: hwachanphy@zjnu.edu.cn} 
	\affiliation{Department of Physics, Zhejiang Normal University, Jinhua 321004, China}
	
	\begin{abstract}
	We consider a bosonic $s$ and $p$ orbital system in a face-centered cubic (FCC) optical lattice, and predict a fluctuation-induced instability towards the orbital analogue of Palmer-Chalker state, which is originally proposed in an electronic spin system. For bosons loaded in the FCC optical lattice, the single-particle spectrum has four degenerate band minima with their crystal momenta forming a tetrahedron in Brillouin zone. In the weakly interacting regime, the ensuing many-particle ground state, at the classical level, underlies a four-sublattice tetrahedral supercell of spontaneously generated $p$-orbital angular momenta through the Bravias-Bloch duality between real and momentum space, and is macroscopically degenerate originating from the geometric frustration. The fluctuations on top of the classical ground state lift its degeneracy and select the Palmer-Chalker ordering of $p$-orbital angular momenta as the quantum ground state through order-by-disorder mechanism. These findings raise the exciting possibility of simulating the Palmer-Chalker state with its orbital counterpart in ultracold atomic gases. 
	\end{abstract}
	
	\date{\today}
	
	\maketitle
	
	\section{Introduction}
	Geometric frustration, usually characterized by a macroscopic degeneracy of classical ground-state manifold, is the key attribute of emerging exotic quantum ground states. Such a degeneracy in the energy landscape arises from competing interactions and can be lifted by quantum fluctuations or additional interactions, thereby selecting a unique ground state. A pertinent issue of relevance is the quantum magnets on geometrically frustrated lattices~\cite{Lacroix11}. For example, the classical ground state of the Heisenberg antiferromagnet on a pyrochlore lattice composed of frustrated corner-sharing tetrahedra has a a macroscopic degeneracy~\cite{Moessner98,Canals00}. The dipolar interaction, which serves as the perturbation to the isotropic Heisenberg exchange, lifts the classical degeneracy and selects the unique magnetic state, called the Palmer-Chalker (PC) state~\cite{Palmer00}. Over the past decades, intensive experimental efforts are denoted to search for this state in pyrochlore oxides~\cite{Gardner10,Hallas18}. To date, the inelastic neutron scattering measurements show compelling evidences of PC state in several compounds including Gd$_2$Sn$_2$O$_7$, Er$_2$Sn$_2$O$_7$ and Er$_2$Pt$_2$O$_7$~\cite{Hallas17,Petit17,Guitteny13,Wills06}. The microscopic origin of the observed PC state, however, has remained enigmatic for the diverse interactions arising in its proximity to competing phases. 
	
	Ultracold atomic gases, on the other hand, have natural advantages in the quantum simulation of artificial solids in optical lattices~\cite{Bloch12,Bloch08}. In particular, the recent experiments have successfully observed the orbital superfluidity in two-dimensional bipartite optical lattices with the sublattices accommodating $s$ and $p$ orbitals~\cite{Wirth11,Parvis12,Olschlager13,Kock15,Jin19}. It is crucial that the coherence between $p$ orbitals established by the tunnelling process via $s$ orbitals leads to an unexpected long lifetime of atoms in high Bloch bands. Theoretically, these experimental settings have also inspired numerous theoretical proposals to simulate the interacting orbital physics in optical lattices~\cite{Li13,Li14,Liu14,Li16,Chen16,Xu16,Liberto16,Pan19}. 

	Here we propose a concrete protocol to realize the PC state with the nontrivial interplay between lattice geometry and orbital anisotropy. It is worth remarking that the nature of PC state studied in the electronic spin systems is a Mott insulator, in which the electronic charge degree of freedom is frozen by strong correlations. In contrast, the PC state we proposed can be induced by the fluctuations from local Hubbard interactions in a weakly interacting Bose gas. Specifically, we generalize the $sp$ orbital system from a two-dimensional bipartite lattice to a three-dimensional face-centered cubic (FCC) lattice by retaining the essential bipartite ingredient. A remarkable feature is that the band structure has four degenerate energetic minima with the crystal momenta forming a tetrahedron in Brillouin zone. This yields a finite-momentum condensate for weakly interacting bosons. Within Gross-Pitaevskii approximation, the multi-orbital Hubbard interaction breaks time reversal symmetry with spontaneous $p$-orbital angular momenta residing on a geometrically frustrated tetrahedral superlattice. The classical ground state thus has a extensive degeneracy due to the emergent frustration, which prevents the system from choosing a unique ground state. The fluctuations described by the standard Bogoliubov theory are further examined in the ground-state selection. Finally, we show that the PC ordering of $p$-orbital angular momenta is favoured by quantum and thermal fluctuations via order-by-disorder mechanism~\cite{Villain80,Henley89,Diep13,Green18}. The predicted PC state in the superfluid phase is characterized by three linearly dispersing Nambu-Goldstone (NG) modes~\cite{Nambu60,Goldstone61} with one arising from the broken global $U\left(1\right)$ gauge field and the other two degenerate modes being protected by point group symmetries. Our findings extend the search of PC state from strongly correlated solid-state materials to weakly interacting Bose gases.
	
	The remainder of this paper is organized as follows: In Sec.~\ref{sec:spp}, we introduce the optical potential for FCC lattice. The band structure is solved with plane-wave expansions. A four-band tight-binding model is also constructed to capture the low-energy spectrum. In Sec.~\ref{sec:mpp}, the classical ground states in the weak coupling limit are obtained within Gross-Pitaevskii approximation. The quantum and thermal fluctuations are further considered to lift the degeneracy of classical ground states. Finally, we summarise our results and discuss the possible experimental detection in Sec.~\ref{sec:summary}. 
	
	\begin{figure}
		\centering
    	\includegraphics[width=0.48\textwidth]{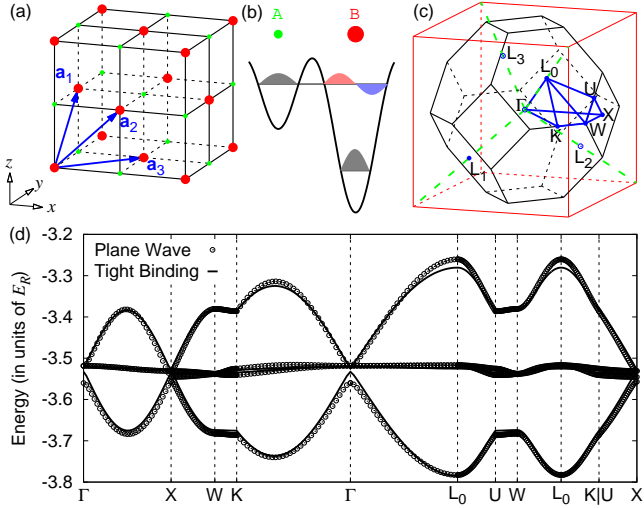}
		\caption{
			(a) FCC Bravais lattice with primitive lattice vectors.
			(b) Schematic plot of the orbital configuration in $\mathcal{A}$ and $\mathcal{B}$ sublattices. 
			(c) Brillouin zone of FCC lattice with high-symmetry lines and points indicated. 
			(d) Band structures with plane-wave expansions (open circles) and tight-binding approximations (solid lines) for the optical potentials $\{V,\Delta V\}=\{-5E_\text{R},-0.9623E_\text{R}\}$ in Eq.~(\ref{eq:OL}) along high symmetry lines indicated in (c). The recoil energy is defined as $E_\text{R}=\pi^2\hbar^2/2md^2$.
		}
		\label{fig:band}
	\end{figure}
	
	\section{Single-Particle Physics}
	\label{sec:spp}
	We start with the optical potential for FCC lattice~\cite{Petsas94,Toader04}
	\begin{equation}
	\mathcal{V}\left(\bm{r}\right)=V\sum_{i=1}^{3}\cos\left[\bar{\bm{b}}_i\cdot\bm{r}\right]
	+\Delta V\sum_{i=0}^{3}\cos\left[\bm{b}_i\cdot\bm{r}\right]
	\label{eq:OL}
	\end{equation}
	where $\bm{b}_1=\pi\left(-\hat{x}+\hat{y}+\hat{z}\right)/d$, 
	$\bm{b}_2=\pi\left(\hat{x}-\hat{y}+\hat{z}\right)/d$,
	and $\bm{b}_3=\pi\left(\hat{x}+\hat{y}-\hat{z}\right)/d$ are the reciprocal lattice vectors with $d$ being the lattice spacing, $\bm{b}_0=-\sum_{i=1}^{3}\bm{b}_i$ and $\bar{\bm{b}}_i=\bm{b}_0+\bm{b}_i$.
	The optical potential in Eq.~(\ref{eq:OL}) produces a three-dimensional bipartite lattice structure with $\Delta V$ dictating the staggered potential difference between $\mathcal{A}$ and $\mathcal{B}$ sublattices shown in Fig.~\ref{fig:band}(a) and \ref{fig:band}(b). Without loss of generality, we will restrict our discussion below with $V<0$ and $\Delta V<0$. As depicted in Fig.~\ref{fig:band}(b), we focus on the case that the $\mathcal{B}$ sublattice is much deeper than the $\mathcal{A}$ sublattice such that the former hosts $\bm{p}\equiv\left(p_x,p_y,p_z\right)$ orbitals, while the latter hosts $s$ orbitals. The low-lying $s$ orbitals in sublattice $\mathcal{B}$ are well separated from the $\bm{p}$ orbitals in energy and thus can be safely neglected. The local minima of the optical potential in Eq.~(\ref{eq:OL}) underlies a FCC Bravais lattice depicted in Fig.~\ref{fig:band}(a). The band structure is firstly solved by the plane-wave expansions. See Appendix~\ref{app:PW} for details. The calculated band structure is shown in Fig.~\ref{fig:band}(d) and has degenerate band minima at four distinct momenta $\bm{L}_{i=\{0,1,2,3\}}=-\bm{b}_i/2$ in the first Brillouin zone. 
	
	To facilitate our understanding, the tight-binding model is desired to capture the band dispersion from plane-wave expansions. The minimal tight-binding model consists of the $\sigma$ bondings $t_{\alpha\beta\sigma}$ between $\alpha$ and $\beta$ orbitals up to the second nearest neighboring sites as well as the on-site energies of $s$ and $\bm{p}$ orbitals. The $\pi$ bonding $t_{pp\pi}$ lies in the nodal plane of the $p$ orbitals and is typically much weaker than the $\sigma$ bonding $t_{pp\sigma}$. Thus, for practical consideration, we neglect the $\pi$ bonding in the tight-binding approximation. We next address the on-site energies of the $s$ and $\bm{p}$ orbitals from symmetry aspects~\cite{Chen18}. Switching to spherical coordinates, the optical potential in Eq.(\ref{eq:OL}) around $\mathcal{A}$ ($-$) and $\mathcal{B}$ ($+$) sublattices can be expressed as
	\begin{eqnarray}
	\mathcal{V}\left(\bm{r}\right) &=& \sum_{\ell \text{ even}} \sum_{m=-\ell}^{\ell} 4\pi i^\ell v^\pm_{\ell m}\left(r\right)Y_{\ell m}\left(\hat{\bm{r}}\right), \label{eq:vexp}\\
	v^\pm_{\ell m}\left(r\right) &\equiv& \sum_{i=1}^{3} Vj_\ell \left(\bar{b}_ir\right)Y^*_{\ell m}\left(\hat{\bar{\bm{b}}}_i\right) \nonumber\\
	&\pm&\sum_{i=0}^{3} \Delta Vj_\ell \left(b_ir\right)Y^*_{\ell m}\left(\hat{\bm{b}}_i\right), \nonumber
	\end{eqnarray}
	where $j_\ell\left(z\right)$ is the spherical Bessel function of the first kind and $Y_{\ell m}\left(\hat{\bm{r}}\right)$ is the spherical harmonic function. The $s$ orbital has zero angular momentum and only receives nonzero correction from the isotropic channel $\ell=0$, which is predicted by the section rule on the orbital angular momentum~\cite{Chen18}. In contrast, the $\bm{p}$ orbitals form an $\ell=1$ angular momentum and thus receive possible corrections from $\ell=0,2$ channels. Direct evaluations show that $v^{+}_{\ell=2,m}\left(r\right)$ in Eq.~(\ref{eq:vexp}) vanishes. Therefore, the $\bm{p}$ orbitals receive nonzero corrections only from the isotropic channel $\ell=0$ and therefore remain degenerate. With these in mind, we denote the on-site energies of $s$ and $\bm{p}$ orbitals as $\epsilon_s$ and $\epsilon_p$, respectively. Introducing a spinor representation for the Bloch field operator $\psi_{\bm{k}}=\left(s_{\bm{k}},p_{x\bm{k}},p_{y\bm{k}},p_{z\bm{k}}\right)^\text{T}$, the tight-binding model in momentum space is then given by 
	\begin{equation}
	H_\text{TB}=\sum_{\bm{k}}\psi_{\bm{k}}^\dagger\mathcal{H}_{\bm{k}}\psi_{\bm{k}} 
	\label{eq:TB}
	\end{equation} 
	where 
	\begin{equation}
	\mathcal{H}_{\bm{k}}=
	\left(
	\begin{matrix}
	\xi_s				&2it_{sp\sigma}s_x		&2it_{sp\sigma}s_y		&2it_{sp\sigma}s_z		\\
	-2it_{sp\sigma}s_x	&\xi_x					&-2t_{pp\sigma}s_xs_y	&-2t_{pp\sigma}s_zs_x	\\ 		
	-2it_{sp\sigma}s_y	&-2t_{pp\sigma}s_xs_y	&\xi_y					&-2t_{pp\sigma}s_ys_z	\\
	-2it_{sp\sigma}s_z	&-2t_{pp\sigma}s_zs_x	&-2t_{pp\sigma}s_ys_z	&\xi_z					
	\end{matrix}
	\right) \nonumber
	\label{eq:TBk}
	\end{equation}
	with
	$s_\mu \equiv \sin k_\mu$, $\xi_s\equiv \sum_{\mu \nu}^{\mu\ne\nu} t_{ss\sigma} \cos k_\mu \cos k_\nu +\epsilon_s$,
	and $\xi_\mu \equiv \sum_{\nu\ne\mu} 2 t_{pp\sigma} \cos k_\mu \cos k_\nu+\epsilon_p$.	It is worth noticing that the relative difference, $\epsilon_s-\epsilon_p$, between the on-site energies of $s$ and $\bm{p}$ orbitals can be continuously tuned through the optical potential $\Delta V$ in Eq.~(\ref{eq:OL}). In the band fitting procedure, the optical potentials in units of recoil energy $E_\text{R}=\pi^2\hbar^2/2md^2$ are chosen as $\{V,\Delta V\}=\{-5E_\text{R},-0.9623E_\text{R}\}$ such that the on-site energies of $s$ and $\bm{p}$ orbitals are degenerate $\epsilon_s=\epsilon_p\equiv\epsilon$, which greatly simplifies our analysis below. Here, we emphasize that the physics we discussed does not sensitively depends on the parameters. As shown in Fig.~\ref{fig:band}(d), the fitted tight-binding model well produces the overall band dispersion from the plane-wave expansions, and faithfully captures the low-energy behavior around the band minima~\cite{TB}. With the constructed tight-binding model, the quasiparticles of the degenerate band minima have energy $\epsilon_{\bm{L}}=-2\sqrt{3t_{sp\sigma}^{2}+t_{pp\sigma}^{2}}-2t_{pp\sigma}+\epsilon$, and are given by  
	\begin{equation}
	\psi_{\bm{L}_i}^\dagger = \cos\Phi s^\dagger_{\bm{L}_i}+i\sin\Phi\hat{\bm{L}}_i\cdot\bm{p}^\dagger_{\bm{L}_i}\text{, } i=\{0,1,2,3\},
	\label{eq:psiL}
	\end{equation} 
	where $\Phi=\arctan \Upsilon$ with the auxiliary function $\Upsilon\equiv\left(t_{pp\sigma}+\sqrt{t^{2}_{pp\sigma}+3t^{2}_{sp\sigma}}\right)/\sqrt{3}t_{sp\sigma}$.
	Based on these quasiparticles, a set of degenerate single-particle states that equally minimize the kinetic energy can be constructed by linear superposition of the band minima $\psi^\dagger=\sum_{i=0}^{3}\phi_i\psi^\dagger_{\bm{L}_i}$. Its manifold lives on the surface $S^7$ in $\mathbb{R}^8$, $\left|\bm{\phi}\right|=1$ with $\bm{\phi}\equiv\left(\phi_0,\phi_1,\phi_2,\phi_3\right)$. Because of the infinite degeneracy of the single-particle states, free bosons cannot condense.	
	
	\section{Weak Interacting Many-particle Physics}
	\label{sec:mpp}

	\subsection{Classical Ground State}	
	
	\begin{figure*}
	\centering
	\includegraphics[width=0.98\textwidth]{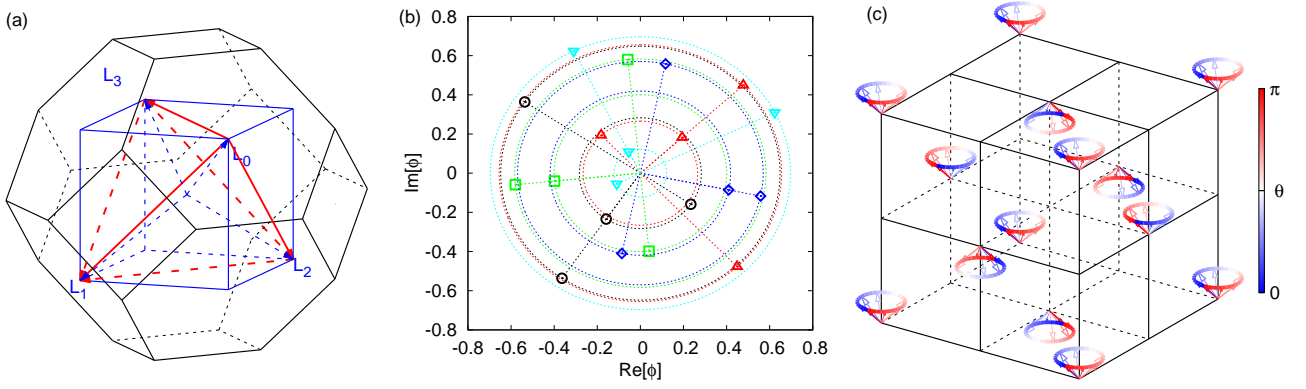}
	\caption{
		(a) The inner blue cube denotes the magnetic Brillouin zone of $p$-orbital angular momenta in (c) due to the interference of condensates at momenta $\bm{L}_{\{0,1,2,3\}}$. 			
		(b) Illustrative the ground-state configurations $\bm{\phi}=\left(\phi_0,\phi_1,\phi_2,\phi_3\right)$ obtained by the imaginary time evolution of Gross-Pitaevskii equation. Different initial configurations in the numerical simulations are indicated by point types. (c) Trajectories of spontaneous $p$-orbital angular momenta with the color encoding the angle $\theta$ in Eq.~(\ref{eq:mfcoef}). 
	}
	\label{fig:itp}
	\end{figure*}	
	
	Having established the single-particle physics, we are then in a position to study how the many-body interaction lifts the  infinite degeneracy of single-particle states. The on-site Hubbard interactions can be experimentally realized through the $s$-wave Feshbach resonance~\cite{Chin10}, and takes the following form
	\begin{eqnarray}
	\hspace{-4mm}
	H_\text{I}=\frac{U_s}{2}\sum_{\bm{r}}\hat{n}_{s\bm{r}}\left(\hat{n}_{s\bm{r}}-1\right)
	+\frac{U_p}{2}\sum_{\bm{r}}\left[\hat{n}^2_{p\bm{r}}-\frac{\hat{n}_{p\bm{r}}+\hat{\bm{J}}^2_{\bm{r}}}{3}\right]
	\label{eq:int}
	\end{eqnarray}
	where $\hat{n}_{s\bm{r}}=s^\dagger_{\bm{r}}s_{\bm{r}}$ and $\hat{n}_{p\bm{r}}=\sum_{\mu}p^\dagger_{\mu\bm{r}}p_{\mu\bm{r}}$ are the density operators for $s$ and $\bm{p}$ orbitals respectively, the $\mu$-component orbital angular momentum operator $\hat{J}^\mu_{\bm{r}}=-i\sum_{\nu\lambda}\epsilon_{\mu\nu\lambda}p^\dagger_{\nu\bm{r}} p_{\lambda\bm{r}}$ and $\epsilon_{\mu\nu\lambda}$ is the Levi-Civita symbol. The interaction parameters can be estimated as $U_s=\frac{4}{3}\left[1-2\Delta V/\left(V+\Delta V\right)\right]^{3/2}U_p\equiv U$ from the harmonic approximation~\cite{Isacsson05,Liu06}. The last term in Eq.~(\ref{eq:int}) enjoys SU($2$) rotational symmetry and favours spontaneous orbital angular momenta to lower the energy, which is analogous to the Hund's coupling for electrons in an atom. According to the orbital Hund's rule, the bosons are not subject to the Pauli exclusion principle and tend to condensate in a single orbital to maximize the orbital angular momentum. 

	We next turn to the many-particle wave function of the condensates. It is worth to mention that the fragmented condensate violates the orbital Hund's rule and does not optimize the orbital Hund's coupling due to the exchange correlations~\cite{Nozieres95}. Below, we shall consider the coherent condensate 
	$\left|\Psi\right>=\frac{1}{\sqrt{\mathcal{N}_0!}}
	\left(\sum_{i=0}^{3}\phi_i\psi_{\bm{L}_i}^\dagger\right)^{\mathcal{N}_0}\left|0\right>,$
	where $\left|0\right>$ denotes the vacuum state and $\mathcal{N}_0$ is the number of condensed bosons.
	Using the coherent state, the time-dependent Gross-Pitaevskii equation can be readily derived through the Euler-Lagrange equation~\cite{Pethick08}
	\begin{equation}
	\frac{\partial \mathcal{L}}{\partial \phi^*_i}
	-\frac{d}{dt}\left(\frac{\partial \mathcal{L}}{\partial \dot{\phi^*_i}}\right)=0\text{, }
	i=\{0,1,2,3\},
	\label{eq:EL}
	\end{equation}
	where the Lagrangian  $\mathcal{L}\equiv\sum_{i=0}^{3}i\frac{\hbar}{2}\left(\phi^*_i\dot{\phi}_i-\phi_i\dot{\phi}^*_i\right)
	-\mathcal{E}\left(\bm{\phi}^*,\bm{\phi}\right)$
	with $\mathcal{E}\left(\bm{\phi}^*,\bm{\phi}\right)\equiv \left<\Psi\right| H_\text{TB}+H_\text{I}\left|\Psi\right>/\mathcal{N}_\text{L}$ the energy functional and $\mathcal{N}_\text{L}$ the number of lattice sites. The derivation of Gross-Pitaevskii equation is presented in Appendix~\ref{app:GPE}. Here let us briefly discuss the symmetry. As shown in Fig.~\ref{fig:itp}(a), $\bm{L}_{0,1,2,3}$ connect the center of a tetrahedron to its vertices. The Lagrangian $\mathcal{L}$ naturally inherits the $T_d$ point group symmetry of the tetrahedron through the single-particle states in Eq.~(\ref{eq:psiL}). The Gross-Pitaevskii equation is numerically solved with the imaginary time evolution by propagating an initial trial state~\cite{Dalfovo99}. With different initial states, a series of degenerate ground states are obtained. We illustrate the ground-state configuration $\bm{\phi}$ by several sets of numerical solutions depicted in Fig.~\ref{fig:itp}(b). The configuration $\bm{\phi}$ splits into two pairs each with identical complex modulus. The complex phases of $\bm{\phi}$ also have internal structures: two $\phi$s have identical phase and the others are $\pi/2$ ahead and behind. With these insights, the numerical solutions can be described by the following analytic expression (up to a global $U\left(1\right)$ phase)
	\begin{equation}
	\bm{\phi}=\frac{1}{\sqrt{2}}\left(i\cos\theta,\cos\theta,-i\sin\theta,\sin\theta\right),0\le\theta<\pi
	\label{eq:mfcoef}
	\end{equation}
	or its counterparts with permutations, which is a manifestation of $T_d$ point group symmetry~\cite{Dresselhaus08}. Moreover, we have also verified that the ground-state energy density in numerical simulations is consistent with the analytic result $\mathcal{E}_0=\epsilon_{\bm{L}}n_0+\left(\cos^4\Phi U_s+ 19\sin^4\Phi U_p/27\right)n_0^2/2$ where the condensation density $n_0=\mathcal{N}_0/\mathcal{N}_\text{L}$. The ground states spontaneously break the time-reversal symmetry and support $p$-orbital angular momenta due to the aforementioned orbital Hund's coupling. The $p$-orbital angular momenta involve the interference between the band minima at $\bm{L}_{\{0,1,2,3\}}$, resulting in a reduced Brillouin zone in Fig.~\ref{fig:itp}(a). With the analytic configuration in Eq.~(\ref{eq:mfcoef}), the trajectories of $p$-orbital angular momenta $\bm{J}_{\bm{r}}$ plotted in Fig.~\ref{fig:itp}(c) have an enlarged unit cell with four sublattice forming a tetrahedron, which we also have confirmed in the numerical simulations. The classical solution of the Gross-Pitaevskii equation partially lifts the single-particle degeneracy on surface $S^7$ and still enjoys the infinite degeneracy arising from the global $U\left(1\right)$ phase and the continuous symmetry characterized by $\theta$ in Eq.~(\ref{eq:mfcoef}). The classical ground-state degeneracy is a consequence of geometric frustration, {\it i.e.} inability to simultaneously minimizes the energy of all bonds in the supperlattice composed of emergent tetrahedra.

	\subsection{Quantum Ground State}
	
	\begin{figure}[t]
	\centering
	\includegraphics[width=0.48\textwidth]{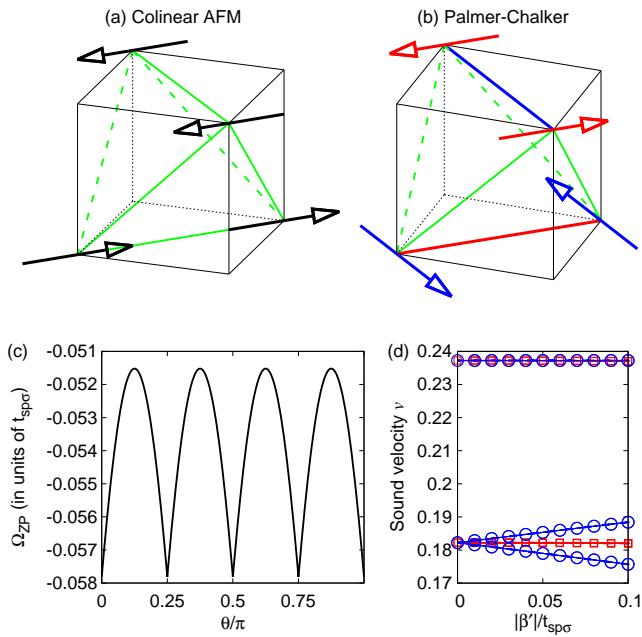}
	\caption{
		Schematic plot of the ordering pattern of $p$-orbital angular momenta on a single tetrahedron for (a) collinear antiferromagnetic ordering at $\theta=0,\pi/2$ and (b) Palmer-Chalker ordering at $\theta=\pi/4,3\pi/4$ (The $p$-orbital angular momenta are in the same color with the parallel edges of tetrahedron). 
		(c) The thermodynamic potential from the zero-point fluctuation $\Omega_{\text{ZP}}$ in the long-wavelength limit with $\{\beta,\beta^\prime,U,n\}=\{t_{sp\sigma}/\sqrt{3},0,t_{sp\sigma}/10,1\}$. (d) Numerical evaluation on the solid-angle averaged sound velocities of Nambu-Goldstone modes for the collinear antiferromagnetic (open red squares) and Palmer-Chalker (open blue circles) states with $\{\beta,U,n\}=\{t_{sp\sigma}/\sqrt{3},t_{sp\sigma}/10,1\}$. For the isotropic case $\beta^\prime=0$, we have also numerically confirmed that the sound velocities of  Nambu-Goldstone modes are $\theta$-independent.  	
	}
	\label{fig:fluct}
	\end{figure}	
	
	The ground states of Gross-Pitaevskii equation can evolve in the continuous-symmetry space without energy cost, {\it e.g.} the trajectories in Fig.~\ref{fig:itp}(c), which makes the system particularly susceptible to fluctuations. We therefore proceed to examine the effects of fluctuations on the degenerated classical ground states. Following the standard Bogoliubov approximation for a weakly interacting Bose gas~\cite{Bogolyubov47,Abrikosov63}, the bosonic field $\psi_{\bm{L}_i+\bm{k}}$ around the band minima $\bm{L}_i$ is separated into the classical condensation $\phi_i$ and the fluctuating field $\phi_{i\bm{k}}$ as $\psi_{\bm{L}_i+\bm{k}}=\phi_i+\phi_{i\bm{k}}$.
	We further project both the kinetic and interaction terms into four lowest branches at band minima, which captures the essential low-energy physics in the long-wavelength limit. 
	Within the formalism of functional integral, the partition function is given by
	\begin{eqnarray}
	\mathcal{Z}=\int\mathcal{D}\left[\bar{\bm{\phi}},\bm{\phi}\right]
	e^{-\mathcal{S}_{\text{eff}}\left[\bar{\bm{\phi}},\bm{\phi}\right]}, 
	\label{eq:pi}
	\end{eqnarray}
	where the effective action up to quadratic order reads
	\begin{eqnarray}
	&&\mathcal{S}_\text{eff}\left[\bar{\bm{\phi}},\bm{\phi}\right]=
	\frac{1}{2}\sum_{k}
	\left(
	\begin{matrix}
	\bar{\bm{\phi}}_{k} &
	\bm{\phi}_{-k} 
	\end{matrix}
	\right) 
	\mathcal{G}_{k}^{-1}
	\left(
	\begin{matrix}
	\bm{\phi}_{k} \\
	\bar{\bm{\phi}}_{-k}
	\end{matrix}
	\right)
	-\sum_{\bm{k}}\frac{\text{Tr}\mathcal{G}^{-1}_{\bm{k},0}}{4},\nonumber\\
	&&\mathcal{G}^{-1}_{k}=
	\left(
	\begin{matrix}
	-i\omega_n+\mathbf{\epsilon}_{\bm{k}}+\mathbf{\Sigma}&
	\mathbf{\Delta} \\
	\mathbf{\Delta}^\dagger &
	i\omega_n+\mathbf{\epsilon}_{-\bm{k}}+\mathbf{\Sigma}^{\text{T}}
	\end{matrix}
	\right)\hspace{-0.1cm}, 	
	k\equiv\left(\bm{k},\omega_n\right).\nonumber
	\end{eqnarray}
	See Appendix~\ref{app:action} for the derivation.
	Here we have defined the auxiliary matrices
	\begin{eqnarray}	
	\mathbf{\Sigma}&=&\left\{\mathcal{U}_s+11\mathcal{U}_p\right\}\mathbf{1}
	+8\mathcal{U}_p\left\{\cos\left[2\theta\right]\gamma^0+i\sin\left[2\theta\right]\sigma^{03}\right\},\nonumber\\
	\mathbf{\Delta}&=&\mathcal{U}_s\left\{\sin\left[2\theta\right]\gamma^5+i\cos\left[2\theta\right]\sigma^{23}\right\}
	+\mathcal{U}_p\left\{4i\gamma^1\gamma^5-4\sigma^{12}\right.\nonumber\\
	&+&\sin\left[2\theta\right]\left(7\gamma^5-4\gamma^2\right)
	+\left.\cos\left[2\theta\right]\left(7\sigma^{23}-4\gamma^3\gamma^5\right)\right\}\nonumber
	\end{eqnarray}
	with gamma matrices $\gamma^{0,1,2,3,5}$ in the Pauli-Dirac representation, $\sigma^{\mu\nu}=\frac{i}{2}\left[\gamma^\mu,\gamma^\nu\right]$, the effective interaction parameters $\{\mathcal{U}_s,\mathcal{U}_p\}=\{U_sn\cos^4\Phi,U_pn\sin^4\Phi/27\}$ and the total density of bosons $n$.
	The band dispersions $\mathbf{\epsilon}_{\bm{k}}=\text{diag}\left(\epsilon_{0\bm{k}},\epsilon_{1\bm{k}},\epsilon_{2\bm{k}},\epsilon_{3\bm{k}}\right)$ are expanded in small $\bm{k}$ around the band minima, and are given by
	\begin{eqnarray}
	\epsilon_{i\bm{k}}=
	\beta\left|\bm{k}\right|^2+\beta^\prime\left(k_yk_z,k_zk_x,k_xk_y\right)\cdot\hat{\bm{L}}_i
	+\mathcal{O}(\left|\bm{k}\right|^4),
	\label{eq:bd}
	\end{eqnarray}
	with $\{\beta,\beta^\prime\}
	=\{{t_{sp\sigma}}/{\sqrt{3}}+t_{pp\sigma},-2\left({t_{pp\sigma}}/{\sqrt{3}}+\sqrt{3}t_{ss\sigma}\right)\}$.
	The $\sigma$ bondings $t_{ss\sigma/pp\sigma}$ between second nearest neighbors are relatively weak. 
	In Eq.~(\ref{eq:bd}), we keep up to the linear order in $t_{ss\sigma/pp\sigma}$.
	A remarkable feature of the second term in Eq.~(\ref{eq:bd}) is the anisotropy, which will be discussed later. Below, we shall first focus on the isotropic case $\beta^\prime=0$.
	The excitation spectrum $\omega_{i\bm{k}}$ is determined by the poles of Green function $\text{det}\mathcal{G}^{-1}_{k}=0$ with an inverse Wick rotation $i\omega_n\to\omega_{\bm{k}}$. In the long-wavelength limit, the excitation spectrum for $\theta\ne\mathbb{Z}{\pi}/{4}$ features two linearly dispersing NG modes, arising from the breaking of aforementioned continuous symmetries. In contrast, at $\theta=0,\pi/2$ and $\theta=\pi/4,3\pi/4$, the lattice exhibits a periodic repetition of the collinear antiferromagnetic (CAFM) and PC ordering of $p$-orbital angular momenta shown in Fig.~\ref{fig:fluct}(a) and \ref{fig:fluct}(b), respectively. The CAFM and PC orderings preserve the $\sigma_d$ and $S_4$ symmetry of $T_d$ point group and has an additional NG mode as a consequence of breaking the continuous symmetry of the $\sigma_d$ and $S_4$ counterpart in Eq.~(\ref{eq:mfcoef}), respectively. This result is also supported by our analytical calculations at $\theta=\mathbb{Z}{\pi}/{4}$. The CAFM and PC states share an identical spectrum, two degenerate gapless modes with sound velocity $v=\sqrt{2t_{sp\sigma}\left(\mathcal{U}_s+3\mathcal{U}_p\right)}/^4\hspace{-0.22cm}\sqrt{3}$ and the other one with velocity $v=\sqrt{2t_{sp\sigma}\left(\mathcal{U}_s+19\mathcal{U}_p\right)}/^4\hspace{-0.22cm}\sqrt{3}$. It is worth mentioning that the ordering of $p-$orbital angular momenta is a manifestation of the relative phases between orbitals and they share the same point group symmetry. For arbitrary $\theta$, analytical results are no longer available. The numerical evaluation on the sound velocities of NG modes suggests that the velocities are $\theta$ independent within numerical resolutions. We next turn to discuss the thermodynamic potential $\Omega=-T\ln\mathcal{Z}=\Omega_{\text{ZP}}+\Omega_{\text{T}}$, originating from the zero-point fluctuation $\Omega_{\text{ZP}}=\sum_{\bm{k}}\left(\sum_{i}2\omega_{i\bm{k}}-\text{Tr}\mathcal{G}_{\bm{k},0}^{-1}\right)/4$
	and the thermal fluctuation $\Omega_{\text{T}}=T\sum_{i\bm{k}}\ln\{1-\exp\left[-\beta\omega_{i\bm{k}}\right]\}$.
	Since the effective action is only valid in the long-wavelength limit, the zero-point fluctuation in this limit is shown in Fig~\ref{fig:fluct}(c), and favours both the CAFM and PC states, which further confirms our previous analysis of NG modes from symmetry aspects. At low temperatures, the thermal fluctuation of thermodynamic potential is dominated by NG modes $\omega_{\bm{k}}=v\left|\bm{k}\right|$ with the contributions following a power law behavior $\Omega_{\text{T}}\propto-T^4/v^3$. 
	Both the CAFM and PC states possess three NG modes and thus have a lower thermodynamic potential due to the entropic gain from the zero point motion of NG modes through thermal fluctuations.
	Finally, we would like to discuss the anisotropy induced by the intra-orbital $\sigma$ bondings $t_{ss\sigma/pp\sigma}$ in Eq.~(\ref{eq:bd}).
	The anisotropy $\beta^\prime$ with the present fitting parameters is weak and can, however, be enhanced through tuning the staggered potential $\Delta V$ in Eq.~(\ref{eq:OL}). Hence we release $\beta^\prime$ as a free parameter. The anisotropic sound velocities are numerically evaluated by averaging over the solid angle of $\bm{k}$ and are found to be independent on the sign of $\beta^\prime$. As shown in Fig.~\ref{fig:fluct}(d), the averaged sound velocities of the degenerate NG modes in PC state are split into two branches with the low-lying one well below those in CAFM state. 
	The Bogoliubov excitation corresponds to simultaneous creation or annihilation of two bosons in the excited states with the momentum $\pm \bm{k}$ relation to the band minima. The band anisotropy in Eq. (9) is characterized by the crystal momentum $\hat{\bm{L}}_i$ around the band minimum $\bm{L}_i$. One distinctive feature between the CAFM and PC states is that the PC state involves the Bose condensation at four band minima and the CAFM state involves only two of four band minima. For the PC state, the Bogoliubov spectrum corresponds to the excitation on top of the Bose condensation at four crystal momenta $\bm{L}_i$ and naturally inherits the band anisotropy. By averaging over the solid angle of $\bm{k}$, the PC state has a lower branch of NG modes than the CAFM state and is selected as the quantum ground state through thermal fluctuations. 
	
	\section{Summary and Discussion}
	\label{sec:summary}
	For weakly interacting Bose gases, we have predicted that the PC instability can be induced by both quantum and thermal fluctuations in a FCC optical lattice. Remarkably, our weak-coupling approach combined with numerical and analytical calculations is also applicable to a generic bosonic superfluidity with multiple degenerate minima in its single-particle spectrum. We would like to briefly discuss the experimental detection. Based on the current experimental techniques in ultracold atoms, the momentum distribution of Bose condensates can be measured by the time-of-flight technique with a sudden expansion of the trapped condensates~\cite{Kastberg95,Greiner01,Pedri01,Kohl05}. The crystal momentum distribution in the first Brillouin zone can be constructed based on the bare momentum distribution~\cite{Song19}. The PC state is uniquely characterized by the identical intensity at the corners of magnetic Brillouin Zone in Fig.~\ref{fig:itp}(a). In addition, the NG modes, as the hallmark of continuous-symmetry breaking, can be directly detected by momentum-resolved Bragg spectroscopy~\cite{Ernst10}. We hope that our work will open new avenues to simulate the strongly correlated magnetic states in solid-state materials using weakly interacting Bose gases in optical lattices.   
	
	\section*{Acknowledgement}
	We thank Xiaopeng Li for helpful discussions. 
	This work is supported by National Natural Science Foundation of China under Grants No. 11704338 and No. 11835011.

	\appendix	
\section{Band Structure Calculation with Plane-Wave Expansions}
\label{app:PW}

In this section, we solve the band structure of the following Hamiltonian
\begin{equation}
\hat{\mathcal H}_\text{OL} =
{\int}d\bm{r}\hat{\Psi}^{\dagger}(\bm{r})\Bigl[-\dfrac{\hbar^{2}}{2m}\bm{\nabla}^{2}
+\mathcal{V}(\bm{r})\Bigr]\hat{\Psi}(\bm{r})
\label{eq:HOL-S}
\end{equation}
with plane-wave expansions~\cite{Ashcroft76}.
For FCC optical lattice, the optical potential $\mathcal{V}\left(\bm{r}+\bm{R}\right)=\mathcal{V}\left(\bm{r}\right)$ is invariant under discrete translation vectors $\bm{R}=N_1\bm{a}_1+N_2\bm{a}_2+N_3\bm{a}_3$ with integer multiples of three primitive vectors 
\begin{eqnarray}
\bm{a}_1 = d\left(\hat{y}+\hat{z}\right),
\bm{a}_2 = d\left(\hat{z}+\hat{x}\right), 
\bm{a}_3 = d\left(\hat{x}+\hat{y}\right) 
\end{eqnarray} 
where $d$ is the lattice spacing.
Making use of Bloch's theorem, the field operator in Eq.~(\ref{eq:HOL-S}) is represented as
$\hat{\Psi}\left(\bm{r}\right)=\sum_{n\bm{k}}\psi_{n\bm{k}}\left(\bm{r}\right)\hat{\Psi}_{n\bm{k}}$ with the Bloch wave function
\begin{equation}
\psi_{n\bm{k}}\left(\bm{r}\right) \equiv
\exp\left[i\bm{k}\cdot\bm{r}\right]
\phi_{n\bm{k}}\left(\bm{r}\right),
\label{eq:bloch-S}
\end{equation}
where the quantum number $n$ is the band index and
the crystal momentum $\bm{k}$ can be composed from the reciprocal lattice vectors with fractional coefficients
\begin{equation}
\bm{k} = k_1\bm{b}_1+k_2\bm{b}_2+k_3\bm{b}_3, 
\end{equation}
and the reciprocal lattice vectors
\begin{eqnarray}
\bm{b}_1 &=& \frac{\pi}{d}\left(-\hat{x}+\hat{y}+\hat{z}\right), \nonumber\\
\bm{b}_2 &=& \frac{\pi}{d}\left(\hat{x}-\hat{y}+\hat{z}\right), \nonumber\\
\bm{b}_3 &=& \frac{\pi}{d}\left(\hat{x}+\hat{y}-\hat{z}\right). 
\end{eqnarray}

The Bloch orbitals $\phi_{n\bm{k}}\left(\bm{r}\right)$ in Eq.~(\ref{eq:bloch-S}) inherit the periodicity of the lattice potential,
{\it i.e.} $\phi_{n\bm{k}}\left(\bm{r}+\bm{R}\right)=\phi_{n\bm{k}}\left(\bm{r}\right)$.
Therefore the Bloch wave function can be further rewritten as a linear combination of plane waves
\begin{equation}
\psi_{n\bm{k}}\left(\bm{r}\right)
=\sum_{\bf G} \exp\left[i\left(\bm{k}+\bf{G}\right)\cdot\bm{r}\right]
\phi_{n\bm{k}}^{\bf{G}}
\end{equation}
with plane-wave vectors ${\bf G}=\text{G}_1\bm{b}_1+\text{G}_2\bm{b}_2+\text{G}_3\bm{b}_3$.
Taking the orthogonality of Bloch wave functions,
the eigenstates of the Hamiltonian in Eq.~(\ref{eq:HOL-S}) can be recast to a coupled set of matrix eigenvalue equations
\begin{equation}
4E_\text{R}\left(\bm{k}+\bf{G}\right)^2\phi_{n\bm{k}}^{\bf{G}}
+\sum_{\bf{G}^\prime}V\left(\bf{G}-\bf{G}^\prime\right)
\phi_{n\bm{k}}^{\bf{G}^\prime}
=\epsilon_{n\bm{k}}\phi_{n\bm{k}}^{\bf{G}}
\end{equation}
where the recoil energy $E_\text{R}=\pi^2\hbar^2/2md^2$.
The Fourier transform of lattice potential is given by
\begin{equation}
V\left(\bf G\right) =
\frac{1}{V_\text{uc}} \int_\text{unit cell} d^3\bm{r}
\mathcal{V}\left(\bm{r}\right)
\exp\left[-i{\bf G}\cdot\bm{r}\right] 
\end{equation}
where $V_\text{uc}=\left|\bm{a}_1\cdot\left(\bm{a}_2\times\bm{a}_3\right)\right|$ is the volume of primitive unit cell.	

\section{Derivation of the Time-Dependent Gross-Pitaevskii Equation}
\label{app:GPE}
The time-dependent Gross-Pitaevskii equation is obtained by neglecting the quantum fluctuations of the operators and replacing them by $c$ numbers, which are usually the averages of operators in the ground state. 
We consider the coherent condensed wave function of an ideal Bose gas
\begin{equation}
\left|\Psi\right>=\frac{1}{\sqrt{\mathcal{N}_0!}}
\left(\sum_{i=0}^{3}\phi_i\psi_{\bm{L}_i}^\dagger\right)^{\mathcal{N}_0}\left|0\right>,
\label{eq:cond-S}
\end{equation}
where $\left|0\right>$ denotes the vacuum state and $\mathcal{N}_0$ is the number of condensed bosons.	
The condensed boson density is given by $n_0=\frac{\mathcal{N}_0}{\mathcal{N}_\text{L}}$ with $\mathcal{N}_L$ being the number of lattice sites. The thermodynamic limit is defined by taking the limit $\mathcal{N}_{0}\to\infty$ and $\mathcal{N}_{\text{L}}\to\infty$ with fixed density $n_0$. In the Gross-Pitaevskii approximation, the operators $\psi_{\bm{L}_i}$ are replaced by $c$ numbers
\begin{eqnarray}
\psi_{\bm{L}_i}\to\sqrt{\mathcal{N}_0}\phi_i.
\end{eqnarray}
Accordingly, the operators $s_{\bm{L}_i}$ and $p_{\mu\bm{L}_i}$ are approximated as  
\begin{eqnarray}
s_{\bm{L}_i}&\to&\sqrt{\mathcal{N}_0}\phi_i\cos\Phi,\\
p_{\mu\bm{L}_i}&\to&i\sqrt{{\mathcal{N}_0}}\phi_i\sin\Phi\hat{\bm{L}}_i^\mu.
\end{eqnarray}
After a lengthy but straightforward algebra, the Lagrangian can be written as 
\begin{equation}
\mathcal{L}\equiv\sum_{i}i\frac{\hbar}{2}\left(\phi^*_i\dot{\phi}_i-\phi_i\dot{\phi}^*_i\right)
-\mathcal{E}\left(\bm{\phi}^*,\bm{\phi}\right)
\label{eq:Lag-S}
\end{equation} 
where the energy functional
\begin{eqnarray}
\mathcal{E}\left(\bm{\phi}^*,\bm{\phi}\right)\equiv \frac{1}{\mathcal{N}_\text{L}}
\left[\left<\Psi\right| H_\text{TB}\left|\Psi\right>
+\left<\Psi\right|H_\text{I}\left|\Psi\right>\right],
\label{eq:ef-S}
\end{eqnarray}
with
\begin{widetext}
\begin{eqnarray}
\left<\Psi\right|H_\text{TB}\left|\Psi\right>&=& \mathcal{N}_0\left[-2\left(\sqrt{2t^2_{sp\sigma}+t^2_{pp\sigma}}+t_{pp\sigma}\right)+\epsilon\right]\sum_i \phi^*_i\phi_i
\equiv\mathcal{N}_0\epsilon_{\bm{L}}\sum_i \phi^*_i\phi_i,\\
\left<\Psi\right|H_\text{I}\left|\Psi\right>&=& 
\frac{n_0}{2}\mathcal{N}_0U_s\cos^4\Phi
\left(\sum_{ij}\phi^{*2}_i\phi^2_j+2\sum_{i\ne j}\phi^*_i\phi_i\phi^*_j\phi_j+\sum_{\left(ijkl\right)}\phi^*_i\phi^*_j\phi_k\phi_l\right)\\
&+&\frac{n_0}{6}\mathcal{N}_0U_p\sin^4\Phi
\left(3\sum_{i}\phi^{*2}_i\phi^2_i+\frac{11}{9}\sum_{i\ne j}\phi^{*2}_i\phi^2_j
+\frac{22}{9}\sum_{i\ne j}\phi^{*}_i\phi_i\phi^{*}_j\phi_j
+\frac{1}{3}\sum_{\left(ijkl\right)}\phi^*_i\phi^*_j\phi_k\phi_l\right).
\end{eqnarray}
\end{widetext}
Here $\epsilon_{\bm{L}}\equiv-2\sqrt{2t^2_{sp\sigma}+t^2_{pp\sigma}}-2t_{pp\sigma}+\epsilon$ and ${\left(ijkl\right)}$ denotes all possible permutations of $\left(0,1,2,3\right)$.
Plugging Eq.~(\ref{eq:Lag-S}) into the Euler-Lagrange equation, 
\begin{equation}
\frac{\partial \mathcal{L}}{\partial \phi^*_i}
-\frac{d}{dt}\left(\frac{\partial \mathcal{L}}{\partial \dot{\phi^*_i}}\right)=0
\text{, }i=\{0,1,2,3\},
\label{eq:EL-S}
\end{equation}
yields a set of coupled equations of motion~\cite{Pethick08}
\begin{equation}
i\hbar\dot{\phi}_i=\frac{\partial \mathcal{E}\left(\bm{\phi}^*,\bm{\phi}\right)}{\partial \phi_i^*}
\text{, }i=\{0,1,2,3\}.
\end{equation}
The ground-state solution can be obtained by numerically evolving the imaginary-time equations of motion~\cite{Dalfovo96}.

\section{Derivation of the Effective Action}
\label{app:action}

In this section, we will derive the low-energy effective action around the band minima at $\bm{L}_{0,1,2,3}$, which describes the quadratic fluctuation on top of the classical solution of the Gross-Pitaevskii equation. 	
Following the standard Bogoliubov approximation, the bosonic field $\psi_{\bm{L}_i+\bm{k}}$ around the band minima $\bm{L}_i$ is separated into the classical condensation $\phi_i$ and the fluctuating field $\phi_{i\bm{k}}$ as $\psi_{\bm{L}_i+\bm{k}}=\phi_i+\phi_{i\bm{k}}$. The quadratic fluctuation includes the following three parts.

First, let us discuss the fluctuation arising from the energy functional of classical ground states.
As illustrated in the main text, the configuration of the classical ground state is given by
\begin{eqnarray}
\bm{\phi}=\frac{1}{\sqrt{2}}\left(i\cos\theta,\cos\theta,-i\sin\theta,\sin\theta\right).
\label{eq:mfcoef-S}
\end{eqnarray} 
Substituting Eq.~(\ref{eq:mfcoef-S}) into the energy functional in Eq.~(\ref{eq:ef-S}), we have
\begin{eqnarray}
\hspace{-6mm}\mathcal{E}\left(\bm{\phi}^*,\bm{\phi}\right)=\epsilon_{\bm{L}}n_0
+\frac{1}{2}\left(\cos^4\Phi U_s+\frac{19}{27}\sin^4\Phi U_p\right)n_0^2
\label{eq:ef2-S}
\end{eqnarray}	
where the condensation density $n_0=\mathcal{N}_0/\mathcal{N}_\text{L}$. 
The total bosons consist of condensed bosons in the band minima and excited bosons in the fluctuating fields.
Therefore, the conservation of bosons is given by
\begin{eqnarray}
\mathcal{N}=\mathcal{N}_0+\sum_{i\bm{k}}\phi_{i\bm{k}}^\dagger\phi_{i\bm{k}}.
\label{eq:num-S}
\end{eqnarray}
Substituting Eq.~(\ref{eq:num-S}) into the energy functional in Eq.~(\ref{eq:ef2-S}),
the quadratic fluctuation can be expressed as
\begin{eqnarray}
\mathcal{E}^{(2)}\left(\bm{\phi}^*,\bm{\phi}\right)&=&
-\frac{1}{\mathcal{N}_\text{L}}\left(\sum_{i\bm{k}}\phi_{i\bm{k}}^\dagger\phi_{i\bm{k}}\right) \\
&\times&\left[\epsilon_{\bm{L}}+\left(\cos^4\Phi U_s+\frac{19}{27}\sin^4\Phi U_p\right)n\right] \nonumber
\end{eqnarray}
with $n$ being total boson density. 

Second, let us turn to the fluctuation from tight-binding model. The band structures of the tight-binding model have four degenerate minima $\epsilon_{\bm{L}}=-2\sqrt{2t^2_{sp\sigma}+t^2_{pp\sigma}}-2t_{pp\sigma}+\epsilon$ at momentum $\bm{L}_{0,1,2,3}$. The eigenstates of the band minima at $\bm{L}_i$ are given by 
\begin{equation}
\psi_i=\left[\cos\Phi,i\sin\Phi\hat{\bm{L}}_i^x,i\sin\Phi\hat{\bm{L}}_i^y,i\sin\Phi\hat{\bm{L}}_i^z\right]^\text{T}.
\label{eq:phi-S}
\end{equation} 	
The effective low-energy band dispersions around the band minima are obtained by projecting the the tight-binding model into the eigenstates in Eq.~(\ref{eq:phi-S}), and take the following form
\begin{eqnarray}
E_{i\bm{k}}&=&\left<\psi_i|\mathcal{H}_{\bm{L}_i+\bm{k}}|\psi_i\right> \label{eq:psieff-S}\\
&=&\epsilon_{\bm{L}}
+\beta|\bm{k}|^2+\beta^\prime\left(k_yk_z,k_zk_x,k_xk_y\right)\cdot\hat{\bm{L}}_i+\mathcal{O}\left(|\bm{k}|^4\right)
\nonumber
\end{eqnarray}   
with $\{\beta,\beta^\prime\}
=\{{t_{sp\sigma}}/{\sqrt{3}}+t_{pp\sigma},-2\left({t_{pp\sigma}}/{\sqrt{3}}+\sqrt{3}t_{ss\sigma}\right)\}$. In Eq.~(\ref{eq:psieff-S}) only the linear order in $t_{ss\sigma/pp\sigma}$ is kept. It is worthy to mention that we have verified that the second-order virtual process in which the boson first hops from the lowest band to the upper bands and then hops back to the lowest band contributes in order $|\bm{k}|^4$ and is thus neglected. Therefore, the quadratic fluctuation in tight-binding model is given by
\begin{eqnarray}
H^{(2)}_\text{TB}=E_{i\bm{k}}\sum_{i\bm{k}}\phi_{i\bm{k}}^\dagger\phi_{i\bm{k}}.
\end{eqnarray}

Finally, we shall discuss the fluctuation from the on-site Hubbard interaction.
Let us illustrate the case for $s$ orbitals first. The fluctuation for $s$ orbitals takes the form 
\begin{widetext}
\begin{eqnarray}
H_{\text{I},s}^{(2)}&=&
\frac{U_s}{2} n\cos^2\Phi\sum_{\{i\}\bm{k}}
\phi_{i_1}\phi_{i_2}s^\dagger_{\bm{L}_{i_3}+\bm{k}}s^\dagger_{\bm{L}_{i_4}-\bm{k}}
\delta_{\bm{L}_{i_1}+\bm{L}_{i_2}-\bm{L}_{i_3}-\bm{L}_{i_4},\bf{G}}\\
&+&
\frac{U_s}{2} n\cos^2\Phi\sum_{\{i\}\bm{k}}
\phi_{i_1}^*\phi_{i_2}s^\dagger_{\bm{L}_{i_3}+\bm{k}}s_{\bm{L}_{i_4}+\bm{k}}
\delta_{-\bm{L}_{i_1}+\bm{L}_{i_2}-\bm{L}_{i_3}+\bm{L}_{i_4},\bf{G}}\\
&+&
\frac{U_s}{2} n\cos^2\Phi\sum_{\{i\}\bm{k}}
\phi_{i_1}^*\phi_{i_2}^*s_{\bm{L}_{i_3}+\bm{k}}s_{\bm{L}_{i_4}-\bm{k}}
\delta_{-\bm{L}_{i_1}-\bm{L}_{i_2}+\bm{L}_{i_3}+\bm{L}_{i_4},\bf{G}}.
\end{eqnarray}
Here we have replaced $n_0$ by $n$, which is correct to the order we are calculating.
After projecting into the lowest band $s_{\bm{L}_i+\bm{k}}\approx\cos\Phi\phi_{i\bm{k}}$ and summation over $L_{\{i\}}$, 
a lengthy but straightforward algebra leads to 
\begin{eqnarray}
H_{\text{I},s}^{(2)}&=&
\frac{U_s}{2} n\cos^4\Phi\sum_{\bm{k}}
\left[
\sum_{ij} \phi_{i}\phi_{i}\phi^\dagger_{j\bm{k}}\phi^\dagger_{j-\bm{k}}
+\sum_{i\ne j} 
2\phi_{i}\phi_{j}\phi^\dagger_{i\bm{k}}\phi^\dagger_{j-\bm{k}}
+\sum_{\left(ijkl\right)} \phi_{i}\phi_{j}\phi^\dagger_{k\bm{k}}\phi^\dagger_{l-\bm{k}}
\right]\\
&+&
\frac{U_s}{2} n\cos^2\Phi\sum_{\bm{k}}
\left[
\sum_{i} \phi^\dagger_{i\bm{k}}\phi_{i\bm{k}}
+\sum_{i\ne j}
\left(
\phi_{i}^*\phi_{j}\phi^\dagger_{i\bm{k}}\phi_{j\bm{k}}
+\phi_{i}^*\phi_{j}\phi^\dagger_{j\bm{k}}\phi_{i\bm{k}}
\right)
+\sum_{\left(ijkl\right)} \phi_{i}^*\phi_{j}\phi^\dagger_{k\bm{k}}\phi_{l\bm{k}}
\right]\\
&+&
\frac{U_s}{2} n\cos^2\Phi\sum_{\bm{k}}
\left[
\sum_{ij} \phi_{i}^*\phi_{i}^*\phi_{j\bm{k}}\phi_{j-\bm{k}}
+\sum_{i\ne j}
2\phi_{i}^*\phi_{j}^*\phi_{i\bm{k}}\phi_{j-\bm{k}}
+\sum_{\left(ijkl\right)} \phi_{i}^*\phi_{j}^*\phi_{k\bm{k}}\phi_{l-\bm{k}}
\right]
\end{eqnarray}
where ${\left(ijkl\right)}$ denotes all possible permutations of $\left(0,1,2,3\right)$. Similarly, the fluctuation for $p$ orbitals is given by
\begin{eqnarray}
H_{\text{I},\bm{p}}^{(2)}&=&
\frac{U_p}{6}n\sin^4\Phi\sum_{\bm{k}}
\left[
\sum_{i} \left(\frac{16}{9}\phi_{i}^2+\frac{11}{9}\sum_j\phi_{j}^2\right)\phi^\dagger_{i\bm{k}}\phi^\dagger_{i-\bm{k}}
+\sum_{i\ne j}\frac{22}{9}\phi_i\phi_j\phi^\dagger_{i\bm{k}}\phi^\dagger_{j-\bm{k}}
+\sum_{\left(ijkl\right)}\frac{1}{3}\phi_i\phi_j\phi^\dagger_{k\bm{k}}\phi^\dagger_{l-\bm{k}}
\right]\\
&+&
\frac{2U_p}{3}n\sin^4\Phi\sum_{\bm{k}}
\left[
\sum_{i} \left(\frac{16}{9}|\phi_{i}|^2+\frac{11}{9}\right)\phi^\dagger_{i\bm{k}}\phi_{i\bm{k}}
+\sum_{i\ne j}\frac{11}{9}\left(\phi_i^*\phi_j+\phi_j^*\phi_i\right)\phi^\dagger_{i\bm{k}}\phi_{j\bm{k}}
+\sum_{\left(ijkl\right)}\frac{1}{3}\phi_i^*\phi_j\phi^\dagger_{k\bm{k}}\phi_{l\bm{k}}
\right]\\
&+&
\frac{U_p}{6}n\sin^4\Phi\sum_{\bm{k}}
\left[
\sum_{i} \left(\frac{16}{9}\phi_{i}^{*2}+\frac{11}{9}\sum_{j}\phi_{j}^{*2}\right)\phi_{i\bm{k}}\phi_{i-\bm{k}}
+\sum_{i\ne j}\frac{22}{9}\phi_i^*\phi_j^*\phi_{i\bm{k}}\phi_{j-\bm{k}}
+\sum_{\left(ijkl\right)}\frac{1}{3}\phi_i^*\phi_j^*\phi_{k\bm{k}}\phi_{l-\bm{k}}
\right].
\end{eqnarray}

Collecting the fluctuations discussed above, the Hamiltonian for the quadratic fluctuation is given by
\begin{eqnarray}
H^{(2)}_{\text{QF}}=\mathcal{N}_\text{L}\mathcal{E}^{(2)}\left(\bm{\phi}^*,\bm{\phi}\right)
+H^{(2)}_\text{TB}+H_{\text{I},s}^{(2)}+H_{\text{I},p}^{(2)}.
\end{eqnarray}
It is straightforward to construct the effective action by following Ref.~\cite{Altland10}.
\end{widetext}

\end{document}